\pdfoutput=1
\documentclass[onecolumn]{article}
\usepackage{algorithm,algcompatible,amsmath,graphicx,here,typearea,amsthm,color,cite,authblk}
\typearea{14}
\algnewcommand\INPUT{\item[\textbf{Input:}]}%
\algnewcommand\OUTPUT{\item[\textbf{Output:}]}%
\algnewcommand\RETURN{\item[\textbf{return}]}%

\newcommand{\amp}[0]{{\bf a}}
\newcommand{\estc}[2]{c^{#1}_{#2}}
\newcommand{\ests}[2]{s^{#1}_{#2}}
\newcommand{\estr}[2]{2^{#2}\theta_{#1}}
\newcommand{\Nfirst}[0]{N_{\rm shot}^{1st}}
\newcommand{\Nsecond}[0]{N_{\rm shot}^{2nd}}

\newcommand{\Nshot}[0]{N_{\rm shot}}

\newcommand{\modtwo}[0]{|_{\bmod2\pi}}
\newtheorem{th.}{Theorem}
\newcommand{\figmedium}[3]{
\begin{figure}
 \centering
 \includegraphics[width=130mm,bb=0 0 1920 1400]{#1}
\caption{#2}
\label{#3}
\end{figure}
}
\makeatletter
\def\@maketitle{%
  \newpage
  \null
  \vskip 2em%
  \begin{center}%
  \let \footnote \thanks
    {\Large\bfseries \@title \par}%
    \vskip 1.5em%
    {\normalsize
      \lineskip .5em%
      \begin{tabular}[t]{c}%
        \@author
      \end{tabular}\par}%
    \vskip 1em%
    {\normalsize \@date}%
  \end{center}%
  \par
  \vskip 1.5em}
\makeatother
\begin{document}
\title{Faster Amplitude Estimation}
	\author{Kouhei Nakaji}
	\affil{Department of Applied Physics and Physico-Informatics \& Quantum Computing Center,
Keio University, Hiyoshi 3-14-1, Kohoku, Yokohama, 223-8522, Japan}
	\maketitle
\begin{abstract}
In this paper, we introduce an efficient algorithm for the quantum amplitude estimation task which is tailored for near-term quantum computers. The quantum amplitude estimation is an important problem which has various applications in fields such as quantum chemistry, machine learning, and finance. Because the well-known algorithm for the quantum amplitude estimation using the phase estimation does not work in near-term quantum computers, alternative approaches have been proposed in recent literature. Some of them provide a proof of the upper bound which almost achieves the Heisenberg scaling. However, the constant factor is large and thus the bound is loose. Our contribution in this paper is to provide the algorithm such that the upper bound of query complexity almost achieves the Heisenberg scaling and the constant factor is small. 
\end{abstract} 
\section{Introduction}
The application of near-term quantum computers has been attracting significant interest recently. In the near-term quantum computers, 
the depth of the circuit and the number of qubits are constrained for reducing the noise. Under the constraints, most of the quantum algorithms which realize quantum speed-up in the ideal quantum computers are not available in the near-term quantum computers. The development of algorithms tailored for near-term quantum computers is demanded.

In this paper, we focus on the problem of quantum amplitude estimation in near-term quantum computers. Quantum amplitude estimation is the problem of estimating the value of $\sin\theta$ in the following equation: $\mathcal{A}|0\rangle_{n}|0\rangle= \sin\theta |\tilde{\Psi}_{1}\rangle_n|1\rangle+\cos\theta|\tilde{\Psi}_{0}\rangle_n|0\rangle$. It is well known that the amplitude estimation can be applied to quantum chemistry, finance and machine learning \cite{chemistry-1,chemistry-2,finance-1,finance-2,finance-3,finance-4,ml-1, ml-2, ml-3, ml-4}. In these applications, the cost of execution $\mathcal{A}$ is high, thus how to reduce the number of calling $\mathcal{A}$ while estimating $\theta$ with required accuracy, is the heart of the problem.

The efficient quantum amplitude estimation algorithm which uses the phase estimation has been well documented in previous literature\cite{amplitude-amplification-1, amplitude-amplification-2}. The algorithm achieves Heisenberg scaling, that is, if we demand that the estimation error of $\sin^2\theta$ is within $\epsilon$ with the probability larger than $1/2$, then the query complexity (required number of the call of the operator $\cal{A}$) is $O(1/\epsilon)$. However, the phase estimation requires a lot of noisy gates of two qubits and therefore is not suitable for near-term quantum computers. Thus, it is fair to inquire if there are any quantum amplitude estimation algorithms which work with less resources and still achieve Heisenberg scaling.

Recently, quantum amplitude estimation algorithms without phase estimation have been suggested in some literature \cite{amplitude-estimation-1,simpler-quantum-counting,quantum-approximate-counting,iterative-amplitude-estimation}. Suzuki et al \cite{amplitude-estimation-1} suggests an algorithm which uses maximum likelihood estimation. It shows that the algorithm achieves Heisenberg scaling numerically but there is no rigorous proofs. Wie \cite{simpler-quantum-counting} also studies the problem in the context of quantum counting. Still, it is not rigorously proved that the algorithm achieves Heisenberg scaling. Aaronson et al\cite{quantum-approximate-counting} suggests an algorithm which achieves Heisenberg scaling and give rigorous proof. However, the constant factor proportional to $\frac{1}{\epsilon}\ln\left(\frac{1}{\delta}\right)$ is large where $\delta$ is the probability that the error is less than $\epsilon$. Grinko et al \cite{iterative-amplitude-estimation} also suggests an algorithm which achieves Heisenberg scaling, but the constant factor is still large in the worst case even though it is shown numerically that the constant is smaller in most of the cases.

In this paper, we propose a quantum amplitude estimation algorithm without phase estimation which acheives the Heisenberg scaling and the constant factor proportional to $\frac{1}{\epsilon}\ln\left(\frac{1}{\delta}\right)$ is smaller than previous literature\cite{quantum-approximate-counting,iterative-amplitude-estimation}. The structure of the paper is as follows. In Section~\ref{algorithm}, we discuss our proposed algorithm. The validity of the algorithm is verified by numerical experiments in Section~\ref{experiment}. In Section \ref{discussion}, we conclude with a discussion on future works. The complexity upper bound is proved in the Appendix.

\section{Algorithm}
\label{algorithm}
In this section, we show our proposed algorithm. Before going into the detail, let us summarize the algorithm briefly. 

Similar to the previous literature, we use the quantum amplitude amplification technique\cite{amplitude-amplification-1}. Given the amplitude $\sin\theta$ that we want to estimate, the quantum amplitude amplification enables us to estimate the values of $\cos(2(2m+1)\theta)$ for each non-negative integer $m$ directly by measurements, and the resulting estimation errors of $\cos(2(2m+1)\theta)$ are of the order $O(\sqrt{s})$ with high probability when $s$ measurements are executed for each $m$. 

Our algorithm estimates the values of $\cos(2(2m+1)\theta)$ for each $m=2^{j-1}(j=1\dots\ell)$ iteratively. As in Kitaev'{}s iterative phase estimation \cite{kitaev} (see also \cite{faster-phase-estimation}), if $2(2^{j}+1)\theta\modtwo(j=1\dots\ell)$ are estimated and those estimation errors are within $\sim\pi/2$, then the value of $\theta$ can be iteratively estimated with error $O(1/2^{\ell})$, meaning that the Heisenberg scaling is achieved. However, the value of $2(2^j+1)\theta\modtwo$ is generally not determinable only by the estimate of $\cos(2(2^j+1)\theta)$ because there is ambiguity whether $2(2^j+1)\theta\modtwo\in[0, \pi]$ or $2(2^j+1)\theta\modtwo\in[\pi, 2\pi]$. Our algorithm solves this ambiguity by taking two-stages method. The algorithm is in the first stage when $2(2^{j}+1)\theta < \pi$. In this stage, $2(2^{j}+1)\theta\modtwo$ can be obtained from the estimate of $\cos(2(2^j+1)\theta)$ without ambiguity by using the inverse cosine function. When the estimate of $2(2^{j_0}+1)\theta \sim \pi/2$ at the iteration $j=j_0$, the algorithm moves to the second stage. In the second stage, $2(2^{j}+1)\theta$ might be larger than $\pi$, hence $2(2^{j}+1)\theta\modtwo$ cannot be determined only by the measurements of $\cos(2(2^j+1)\theta)$ because of the above mentioned ambiguity. However, by combining the measurements of $\cos(2(2^j+2^{j_0} + 1)\theta)$ with those of $\cos(2(2^j+1)\theta)$, the value of $\sin(2(2^j+1)\theta)$ can be estimated by using the trigonometric addition formula, and accordingly $2(2^{j}+1)\theta\modtwo$ can be determined without the ambiguity. As a result, the algorithm can estimate the value of $\theta$ with the error less than $O(1/2^{\ell})$.

The algorithm in the reference \cite{iterative-amplitude-estimation} suggests a different approach for solving the ambiguity. However their method requires precise measurements of cosine in the worst case and therefore the complexity upper bound becomes loose. On the other hand, our proposed algorithm works with relatively rough estimates of cosine even in the worst case. Thus, as we will see in Section \ref{subsection-complexity} and Appendix \ref{appendix-upper-bound}, the complexity upper bound of our algorithm is tighter than existing method.

In the following discussion in this section, we first define the problem in Section \ref{preliminary}. Next we look into the detail of the algorithm and finally we show the complexity upper bound of our algorithm.
\subsection{Preliminary}
\label{preliminary}
The quantum amplitude estimation is the problem of estimating the value of $\amp$ in the following equation:
\begin{flalign}
\label{definision}
&|\Psi\rangle\equiv\mathcal{A}|0\rangle_{n}|0\rangle =
\amp|\tilde{\Psi}_{1}\rangle_{n}|1\rangle +
\sqrt{1 - \amp^2}|\tilde{\Psi}_{0}\rangle_n|0\rangle,
\end{flalign}
where $\amp\in [0, 1]$. In applications, it often takes cost to execute $\mathcal{A}$, thus reducing the number of calling $\mathcal{A}$ while estimating $\amp$ with required accuracy is the heart of the problem.

As we see later, our proposed algorithm works correctly if the amplitude is less than or equals to $1/4$. However, imposing the condition on $\amp$ is not necessary because the amplitude can be attenuated by adding an extra ancilla qubit as follows:
\begin{flalign}
\label{R_1/4}
|\Psi^{\prime}\rangle\equiv X |0\rangle_{n}|00\rangle &=
\frac{\amp}{4}|\tilde{\Psi}_{1}\rangle_{n}|11\rangle +
\frac{\sqrt{15}{\amp}}{4}|\tilde{\Psi}_{1}\rangle|10\rangle + 
\frac{\sqrt{1-{\amp}^2}}{4}|\tilde{\Psi}_{0}\rangle|01\rangle + 
\frac{\sqrt{15(1-{\amp}^2)}}{4}|\tilde{\Psi}_{0}\rangle|00\rangle. \nonumber \\
&=\sin\theta|\tilde{\Psi}_{1}\rangle_{n}|11\rangle + 
\cos\theta|{\perp}\rangle, 
\end{flalign}
where $X = \cal{A} \otimes \cal{R}$ and $\cal{R}$ operates as follows:
\begin{flalign}
& R|0\rangle = \frac{1}{4} |1\rangle + \frac{\sqrt{15}}{4} |0\rangle.
\end{flalign}
In the last line of (\ref{R_1/4}), we define $\sin\theta \equiv \amp/4$ and $|\perp\rangle$ as a state orthogonal to $|\tilde{\Psi}_{1}\rangle_{n}|11\rangle$. As expected, the amplitude is attenuated as $\sin\theta \in [0, 1/4]$ and therefore
\begin{flalign}
\label{phi-bound}
&0 \leq \theta < 0.252.
\end{flalign}
Thus, instead of estimating the value of $\amp$ directly, we estimate the value of $\theta$ and convert it to $\amp$.
The condition (\ref{phi-bound}) is utilized as the initial bound in our proposed algorithm. 

Similar to the original amplitude amplification\cite{amplitude-amplification-1}, we define an operator ${\bf Q}$ as
\begin{flalign}
\mathbf{Q} &\equiv X(I_{n + 2} - 2|0\rangle_{n+2}\langle0|_{n+2}) X^{\dag}(I_{n+2} - 2I_n\otimes|11\rangle\langle11|),  
\end{flalign}
where $I_n$ is the identity operator in $n$ dimension. It is worth mentioning that 
\begin{flalign}
\label{multiple-grover}
& \mathbf{Q}^{m}|\Psi^{\prime}\rangle=\sin (2 m+1) \theta|\tilde{\Psi}_{1}\rangle_n|11\rangle +\cos (2 m+1) \theta|\perp\rangle.
\end{flalign}
We get the estimates of $\cos(2(2m + 1)\theta)$ by measuring the state (\ref{multiple-grover}) for multiple $m$; the following defined $\estc{}{m}$ readily computable from the measurement result, is an estimate of $\cos(2(2m + 1)\theta)$:
\begin{flalign}
\label{cos-estimation}
&\estc{}{m} \equiv 1 - 2\frac{N_{11}}{N_{\rm shot}},
\end{flalign}
where $N_{11}$ is the number of the results of the measurements in which the last two qubits in (\ref{multiple-grover}) are both one and $N_{\rm shot}$ is the total number of measurements of the state (\ref{multiple-grover}). The estimation error of $\estc{}{m}$ can be evaluated by using the Chernoff bound for the Bernoulli distribution as discussed in \cite{kitaev}, i.e., given the confidence interval of $\estc{}{m}$ as $[\estc{\rm min}{m}, \estc{\rm max}{m}]$, the bounds of the interval are computed as
\begin{flalign}
\label{chernoff-bound}
\estc{\max}{m} = \min\left[1, \estc{}{m} + \sqrt{\ln\left(\frac{2}{\delta_c}\right)\frac{12}{\Nshot}}\right], \qquad \estc{\min}{m} = \max\left[-1, \estc{}{m} - \sqrt{\ln\left(\frac{2}{\delta_c}\right)\frac{12}{\Nshot}}\right].
\end{flalign}
where $\delta_c$ is the probability that the true value of $\estc{}{m}$ (i.e. $\cos(2(2m + 1)\theta)$) is out of the interval. For later purpose, we define three functions: ${\bf COS}(m, N_{\rm shot})$, ${\bf CHERNOFF}(\estc{}{m}, N_{\rm shot}, \delta_{c})$ and ${\rm atan}(s, c)$. The function ${\bf COS}(m, N_{\rm shot})$ returns $\estc{}{m}$ as a result of $N_{\rm shot}$ times measurements of the state (\ref{multiple-grover}). The function ${\bf CHERNOFF}(\estc{}{m}, N_{\rm shot}, \delta_{c})$ returns the confidence interval $[\estc{\rm min}{m}, \estc{\rm max}{m}]$ of $\estc{}{m}$ computed from the parameters: $\estc{}{m}$, $\Nshot$ and $\delta_c$. The function ${\rm atan}(s, c)$ is an extended arctangent function defined in the realm $c, s \in [-1, 1]$ as 
\begin{flalign}
\label{atan}
{\rm atan}(s, c) = 
\begin{cases}
    \arctan(s/c)&({\rm }c > 0) \\
    \pi/2 &(c = 0, s > 0) \\
    0 &(c = 0, s = 0) \\
    -\pi/2 &(c = 0, s < 0) \\
    \pi + \arctan(s/c)&({c < 0, s \geq 0}) \\
    -\pi + \arctan(s/c)&({c < 0, s < 0}). \\
\end{cases}
\end{flalign}

Finally, we define $N_{\rm orac}$ as the number of calls of $\mathbf{Q}$ required for estimating $\theta$. Our objective in this paper is providing an algorithm to estimate $\theta$ with required accuracy while reducing the number of $N_{\rm orac}$.

\subsection{Proposed Algorithm}
\label{section-algorithm}
In this subsection, we show our proposed algorithm. 
Our procedure is shown in {\bf Algorithm \ref{alg:estimate_rho}}\footnote{The source code of the algorithm is shown in https://github.com/quantum-algorithm/faster-amplitude-estimation}. Given $[\theta_{\min}^j, \theta_{\max}^j]$ as the confidence interval of $\theta$ in $j$-th iteration, the algorithm updates the values of $\theta_{\max}^j$ and $\theta_{\min}^j$ so that $\theta_{\max}^j - \theta_{\min}^j$ becomes smaller in each iteration.
The total iteration count $\ell$ is a parameter given by users of the algorithm and it is chosen so that the final result satisfies the required accuracy. As we see later, given acceptable error of the amplitude $\epsilon$, $\epsilon\sim 1/2^{\ell}$. Therefore, it is suffice to take $\ell$ as $\ell\sim\log_2(1/\epsilon)$.

\begin{algorithm}
  \caption{Faster Amplitude Estimation ($\delta_c$ and $\ell$ as the parameters)}\label{alg:estimate_rho}
  \begin{algorithmic}[1]
  	\STATE \#$\theta^j_{\min}$ and $\theta^j_{\max}$: the confidence interval of $\theta$ in $j$-th iteration.
  	\STATE Set $\theta^0_{\min}$ to $0$ and $\theta^0_{\max}$ to $0.252$.
  	\STATE Set $\Nfirst=1944 \ln \left(\frac{2}{\delta_c}\right)$ and $\Nsecond=972\ln \left(\frac{2}{\delta_c}\right)$.
  	\STATE Set ${\rm FIRST\_STAGE}$ to ${\rm true}$.
  	\STATE Set $j_0$ to $\ell$. 
    \FOR{$j = 1$ to $\ell$}
    \IF{FIRST\_STAGE}
    	\STATE Set $\estc{}{2^{j-1}}$ to $\mathbf{COS}(2^{j-1}, N_{\rm shot}^{\rm 1st})$.
    	\STATE Set $\estc{\min}{2^{j-1}}, \estc{\max}{2^{j-1}}$ to $\mathbf{CHERNOFF}(\estc{}{2^{j-1}}, N_{\rm shot}^{\rm 1st}, \delta_c)$. 
    	\STATE Set $\theta_{\max}^j = \arccos(\estc{\min}{2^{j-1}})/(2^{j+1} + 2)$ and $\theta_{\min}^j =  \arccos(\estc{\max}{2^{j-1}})/(2^{j+1} + 2)$.
    	\IF{$2^{j+1} \theta_{\max}^j \geq \frac{3\pi}{8}$ and $j < \ell$}
    		\STATE Set $j_0$ to $j$.
    		\STATE Set $\nu = 2^{j_0}(\theta_{\max}^{j_0} + \theta_{\min}^{j_0})$ \# the estimate of $2^{j_0 + 1}\theta$  
    		\STATE Set ${\rm FIRST\_STAGE}$ to ${\rm false}$.
    	\ENDIF
    \ELSE
    	\STATE Set $\estc{}{2^{j-1}}$ to $\mathbf{COS}(2^{j-1}, N_{\rm shot}^{\rm 2nd})$.
    	\STATE Set $\ests{}{2^{j-1}}$ to $(\estc{}{2^{j-1}}\cos\nu - \mathbf{COS}(2^{j-1} + 2^{j_0-1}, N_{\rm shot}^{\rm 2nd}))/\sin\nu$.
    	\STATE Set $\rho_j = {\rm atan}\left(\ests{}{2^{j-1}}, \estc{}{2^{j-1}}\right)$. 
		\STATE Set $n_j$ to $[\frac{1}{2\pi}\left((2^{j+1} + 2)\theta_{\max}^{j-1} - \rho_j +\pi/3 \right)]$ where $[x]$ is the largest integer which does not exceed $x$.  
		\STATE Set $\theta_{\min}^j = (2\pi n_j + \rho_j - \pi/3)/(2^{j+1} + 2)$ and $\theta_{\max}^j = (2\pi n_j + \rho_j + \pi/3)/(2^{j+1} + 2)$.
    \ENDIF
    \ENDFOR
    \RETURN $(\theta_{\min}^{\ell} + \theta_{\max}^{\ell})/2$, estimate of $\theta$ where the probability that $\theta\in[\theta_{\min}^j, \theta_{\max}^j]$ is larger than $1-(2\ell - j_0)\delta_c$.
  \end{algorithmic}
\end{algorithm}

Even though $\theta$ is not always inside the confidence interval: $[\theta_{\min}^j, \theta_{\max}^j]$, the probability is bounded and exponentially decreases as $\Nfirst$ and $\Nsecond$ increases. Thus, for simplicity, we discuss only the case when $\theta\in[\theta_{\min}^j, \theta_{\max}^j]$ holds for all $j$s in this subsection. As we show later, the probability that  $\theta\in[\theta_{\min}^j, \theta_{\max}^j]$ holds for all $j$ is larger than $1-2\ell\delta_c$.

In the following, we show how our algorithm works. As we see in {\bf Algorithm \ref{alg:estimate_rho}}, there are two stages and the estimation methods are different in each stage. At the beginning of the iteration ($j = 1$), the algorithm is in the first stage and later the algorithm may change into the second stage if a condition is satisfied. We show the detail in the following.
\subsubsection*{\underline{First Stage}}
The algorithm is in the first stage when $j=1$ or when $j>1$ and all $\estr{\max}{k+1}^k (k=1\dots j-1)$ satisfy $\estr{\max}{k+1}^k < \frac{3\pi}{8}$. In this stage, $\theta_{\min}^j, \theta_{\max}^j$ is gotten by inverting $\estc{\rm min}{2^{j-1}}$ and $\estc{\rm max}{2^{j-1}}$ as 
\begin{flalign}
\label{first-theta}
\theta_{\max}^j &= \frac{\arccos(\estc{\min}{2^{j-1}})}{2^{j+1} + 2},\qquad \theta_{\min}^j =  \frac{\arccos(\estc{\max}{2^{j-1}})}{2^{j+1} + 2}
\end{flalign}
because $(2^{j+1} + 2)\theta < \pi$ is guaranteed as the following argument; if $j = 1$, the bound (\ref{phi-bound}) leads to $(2^{1+1} + 2)\theta < 1.52 < \pi$, and if $j > 1$ and $2^{k+1}\theta_{\max}^k< \frac{3\pi}{8}$ for $(k=1\dots j-1)$ then
\begin{flalign}
& (2^{j+1} + 2)\theta < 2(2^j \theta_{\max}^{j-1}) + 2\theta < 3/4\pi + 0.504 < \pi.
\end{flalign}
The algorithm changes into the second stage  
if $2^{j+1}\theta_{\max}^{j}\geq3\pi/8$. The two values are memorized for the purpose of our utilizing them in the second stage. One is $j_0$ defined as the last value of $j$ in the first stage. Another is $\nu$ defined as
\begin{flalign}
\label{nu}
& \nu = 2^{j_0 + 1} \times \frac{\theta_{\max}^{j_0} + \theta_{\min}^{j_0}}{2}.
\end{flalign}
Note that above defined $\nu$ is an estimate of $2^{j_0+1}\theta$ and the confidence interval is obtainable from the Chernoff bound.

In case that $2^{j + 1}\theta_{\max}^j$ is less than $3\pi/8$ for all ${j}(<\ell)$, the algorithm finishes without going to the second stage and the final result is $(\theta_{\max}^{\ell} + \theta_{\min}^{\ell})/{2}$. The value of $j_0$ is set to $\ell$. In the case, the error of the final result is at most $\Delta\theta\equiv(\theta_{\max}^{\ell} - \theta_{\min}^{\ell})/{2} = (\arccos(\estc{\min}{2^{\ell-1}}) - \arccos(\estc{\max}{2^{\ell-1}}))/(2^{\ell+2}+4)$. Thus, the error of the amplitude is bounded as
\begin{flalign}
\label{amp-error}
\epsilon &= 4\left(\sin(\theta + \Delta\theta) - \sin\theta \right)<4\Delta\theta <\frac{\arccos(\estc{\min}{2^{\ell-1}}) - \arccos(\estc{\max}{2^{\ell-1}})}{2^{\ell}}.
\end{flalign}
The probability that $\theta$ is inside the confidence interval is 
$(1-\delta_c)^{\ell} > 1 - \ell \delta_c (=1-(2\ell - j_0)\delta_c)$.
\subsubsection*{\underline{Second Stage}}
In the second stage, $(2^{j+1} + 2)\theta$ may be larger than $\pi$. Thus, the value of $(2^{j+1} + 2)\theta$ can not be estimated by inverting $\estc{}{2^{j-1}}$. However, it is still possible to estimate the value of $(2^{j+1} + 2)\theta$ by utilizing the results of measurements in other angle: $(2^{j+1}+2^{j_0+1}+2)\theta$, in addition to the bounds of $\theta$ gotten in the previous iteration. Here, firstly we show how to estimate $(2^{j+1} + 2)\theta |_{{\rm mod} 2\pi}$, next we show how to estimate $(2^{j+1} + 2)\theta$ without ${\rm mod}(2\pi)$ ambiguity.
\subsubsection*{\rm\underline{(i)The estimate of $(2^{j+1} + 2)\theta |_{{\rm mod} 2\pi}$}}
To estimate $(2^{j+1} + 2)\theta |_{{\rm mod} 2\pi}$, not only the estimate of $\cos((2^{j+1} + 2)\theta)$ (i.e., $\estc{}{2^{j-1}}$) but also the estimate of $\sin((2^{j+1} + 2)\theta)$ are necessary. The estimate of $\sin((2^{j+1} + 2)\theta)$ is not directly obtainable from measurements but can be computed by the following procedure. From the trigonometric addition formula:
\begin{flalign}
	\cos((2^{j+1}+2^{j_0+1}+2)\theta) = \cos((2^{j+1}+2)\theta)\cos(2^{j_0+1}\theta) - \sin((2^{j+1}+2)\theta)\sin(2^{j_0+1}\theta),
\end{flalign}
if $\sin(2^{j_0+1}\theta)$ is not zero, 
\begin{flalign}
 \sin((2^{j+1}+2)\theta) = 
 \frac{\cos((2^{j+1}+2)\theta)\cos(2^{j_0+1}\theta) - \cos((2^{j+1}+2^{j_0+1}+2)\theta)}{\sin(2^{j_0+1}\theta)}.
\end{flalign}
Replacing $\cos((2^{j+1}+2)\theta$ as $\estc{}{2^{j-1}}$, $2^{j_0+1}\theta$ as $\nu$ and $\cos((2^{j+1}+2^{j_0+1}+2)\theta$ as $\estc{}{2^{j-1}+2^{j_0+1}}$ in the right hand side of the above formula, we can define $\ests{}{2^{j-1}}$ as
\begin{flalign}
\label{p-sin-estimation}
&\ests{}{2^{j - 1}} = \frac{\estc{}{2^{j - 1}}\cos\nu - \estc{}{2^{j-1} + 2^{j_0-1}}}{\sin\nu},
\end{flalign}
then $\ests{}{2^{j-1}}$ becomes the estimate of $\sin((2^{j+1} + 2)\theta)$.
The estimation error of $\ests{}{2^{j - 1}}$ reflects the estimation errors of $\estc{}{2^{j - 1}}$,  $\estc{}{2^{j-1} + 2^{j_0-1}}$ and $\nu$,  
which is discussed in Appendix \ref{appendix-upper-bound}. 
It is straightforward to get the estimate of $(2^{j+1} + 2)\theta |_{{\rm mod} 2\pi}$ from $\ests{}{2^{j-1}}$ and $\estc{}{2^{j-1}}$; if we define $\rho_j$ $\in [-\pi, \pi]$ as
\begin{flalign}
\label{rho-j-estimation-difficult}
&\rho_j = {\rm atan}\left(\ests{}{2^{j-1}}, \estc{}{2^{j-1}}\right),
\end{flalign}
then $\rho_j$ is an estimate of $(2^{j+1} + 2)\theta |_{{\rm mod} 2\pi}$. 

\figmedium{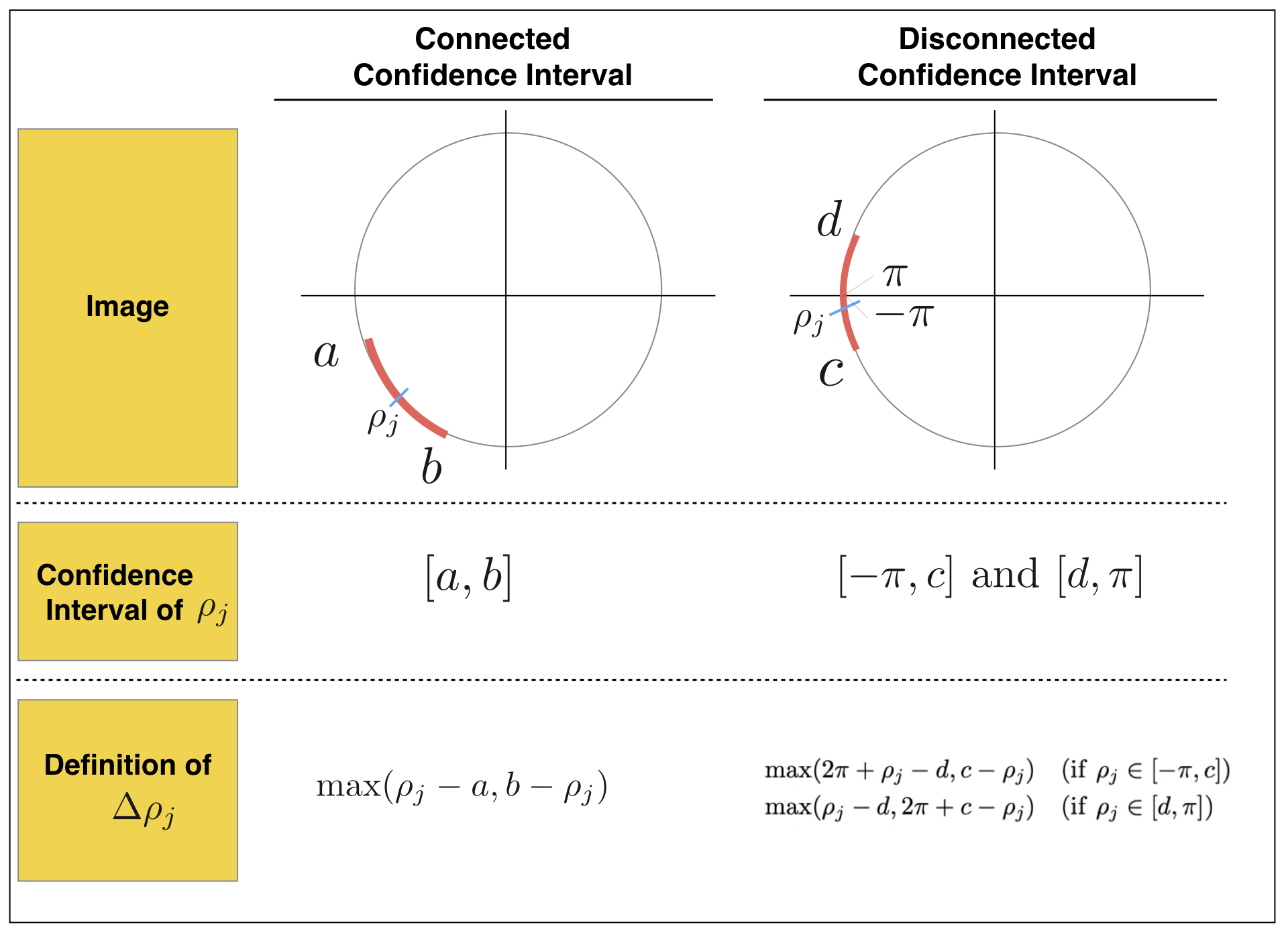}{The overview of the definition of $\Delta \rho_j$.}{definition}

The confidence interval of $\rho_j$ can be derived from those of $\estc{}{2^{j - 1}}$,  $\estc{}{2^{j-1} + 2^{j_0-1}}$ and $\nu$ as in the case of $\ests{}{2^{j-1}}$. Note that there are two types of the confidence interval. One is the connected confidence interval, meaning that there is no discontinuities in the confidence interval, e.g., $[-\pi/3, \pi/4]$. Another is disconnected confidence interval, meaning that the confidence interval is separated to an interval containing $-\pi$ and an interval containing $\pi$, e.g., $[-\pi, -2\pi/3]$ and $[3\pi/4,\pi]$, which is realized when the confidence interval of $\estc{}{2^{j-1}}$ contains $-1$ and that of $\ests{}{2^{j-1}}$ contains 0\footnote{If both the confidence intervals of $\estc{}{2^{j-1}}$ and $\ests{}{2^{j-1}}$ contain $0$, the confidence interval of $\rho_j$ has discontinuity in $\rho_j = \pi/2, -\pi/2$. However, as long as $\Nfirst$ and $\Nsecond$ takes the upper bound value derived in Appendix~\ref{appendix-upper-bound}, the estimation errors are bounded so that either the confidence interval of $\estc{}{2^{j-1}}$ or that of $\ests{}{2^{j-1}}$ does not contain 0 because $(\estc{}{2^{j-1}})^2 + (\ests{}{2^{j-1}})^2 \simeq 1$ holds. Thus, we do not discuss the type of discontinuity in the following.}. In the connected confidence interval case, given interval as $[a, b]$, we define $\Delta\rho_j = \max(\rho_j - a, b - \rho_j)$. On the other hand, in the disconnected confidence interval case, given intervals as $[-\pi, c]$ and $[d, \pi]$, we define $\Delta \rho_j$ as 
\begin{flalign}
\label{delta-rho-disconnected}
\Delta \rho_j = 
\begin{cases}
   \max(2\pi + \rho_j -d, c - \rho_j)&({\rm if\ }\rho_j \in [-\pi, c])\\
   \max(\rho_j -d, 2\pi + c - \rho_j)&({\rm if\ }\rho_j  \in [d, \pi])
\end{cases}.
\end{flalign}
The overview of the definition of $\Delta\rho_j$ is shown in Figure~\ref{definision}.
The above defined $\Delta \rho_j$ can be interpreted as the estimation error of $\rho_j$ in a sense that 
\begin{flalign}
\label{no-mod-formula}
2\pi n_j  + \rho_j - \Delta \rho_j \leq (2^{j+1} + 2)\theta \leq 2\pi n_j + \rho_j + \Delta \rho_j
\end{flalign}
holds with an unknown integer $n_j$ as long as the true value of $\rho_j$ (i.e. $(2^{j+1}+2)\theta|_{\rm \mod 2\pi}$) is inside the confidence interval. 

\subsubsection*{\rm\underline{(ii)The estimate of $(2^{j+1} + 2)\theta$}}
Now let us show how $(2^{j+1} + 2)\theta$ is estimated from $\rho_j$.
Using (\ref{no-mod-formula}) and the inequality,
\begin{flalign}
\label{no-mod-formula-2}
(2^{j+1} + 2)\theta^{j-1}_{\min} \leq (2^{j+1} + 2)\theta \leq (2^{j+1} + 2)\theta^{j-1}_{\max},
\end{flalign}
it can be shown that 
\begin{flalign}
\label{integer-mod-formula}
(2^{j+1} + 2)\theta^{j-1}_{\min} - \rho_j - \Delta\rho_j \leq 2\pi n_j \leq (2^{j+1} + 2)\theta^{j-1}_{\max} - \rho_j + \Delta\rho_j .
\end{flalign}
Thus, if 
\begin{flalign}
\label{theta-formula}
(2^{j+1} + 2)(\theta^{j-1}_{\max} - \theta^{j-1}_{\min}) + 2\Delta\rho_j < 2\pi
\end{flalign}
then $n_j$ can be uniquely determined as 
\begin{flalign}
\label{n-formula}
n_j = \frac{1}{2\pi}[(2^{j+1} + 2)\theta_{\max}^{j-1} - \rho_j + \Delta \rho_j]
\end{flalign}
where $[x]$ is the largest integer that does not exceed $x$. By using (\ref{no-mod-formula-2}) and (\ref{n-formula}), 
it can be inductively shown that if all $\rho_k(k=j_0 + 1 \dots j-1)$ are determined with the precision of $\Delta\rho_k \leq \pi/3$ then the condition (\ref{theta-formula}) is satisfied. 
\vspace{0.5cm}

Although (\ref{no-mod-formula}) with $n_j$ in (\ref{n-formula}) gives the upper/lower bounds of $(2^{j+1}+2)\theta$, a complicated procedure is necessary for evaluating $\Delta \rho_j$ in the algorithm. Thus, in our algorithm, instead of estimating $\Delta \rho_j$, we set the upper/lower bounds of $\theta$ at the $j$-th iteration as
\begin{flalign}
\label{theta-estimation}
\theta_{\min}^{j} = \frac{2\pi n_j + \rho_j - \pi/3}{2^{j+1} + 2}, \qquad \theta_{\max}^{j} = \frac{2\pi n_j+ \rho_j + \pi/3}{2^{j+1} + 2},
\end{flalign}
and 
\begin{flalign}
\label{theta-estimation}
n_j = \frac{1}{2\pi}[(2^{j+1} + 2)\theta_{\max}^{j-1} - \rho_j + \pi/3],
\end{flalign}
which are correct as far as $\Delta \rho_j \leq \pi/3$. In Appendix \ref{appendix-upper-bound}, we show that for all $j(>j_0)$, $\Delta\rho_j\leq\pi/3$ holds and (\ref{no-mod-formula}) is satisfied with the probability larger than $1-(2\ell-j_0)\delta_c$ when at least
\begin{flalign}
\label{n-formula}
\Nfirst=1944 \ln \left(\frac{2}{\delta_c}\right),\qquad\Nsecond=972\ln \left(\frac{2}{\delta_c}\right).
\end{flalign}

In the $\ell$-th iteration, the final result is set to $(\theta_{\max}^{\ell} + \theta_{\min}^{\ell})/2$. Then, the error of the final result $\Delta\theta$ is less than $\Delta\theta=(\theta_{\max}^{\ell} - \theta_{\min}^{\ell})/2 \leq \pi/(3\cdot2^{\ell + 1})$. Thus, the error of the amplitude is
\begin{flalign}
\label{amp-error}
\epsilon &= 4\left(\sin(\theta + \Delta\theta) - \sin\theta \right)<4\Delta\theta <\frac{\pi}{3\cdot 2^{\ell-1}}. 
\end{flalign}
We show the overview of our algorithm when $\ell=5$ and $j_0=3$ in Figure~\ref{summary-of-algorithm}.
\figmedium{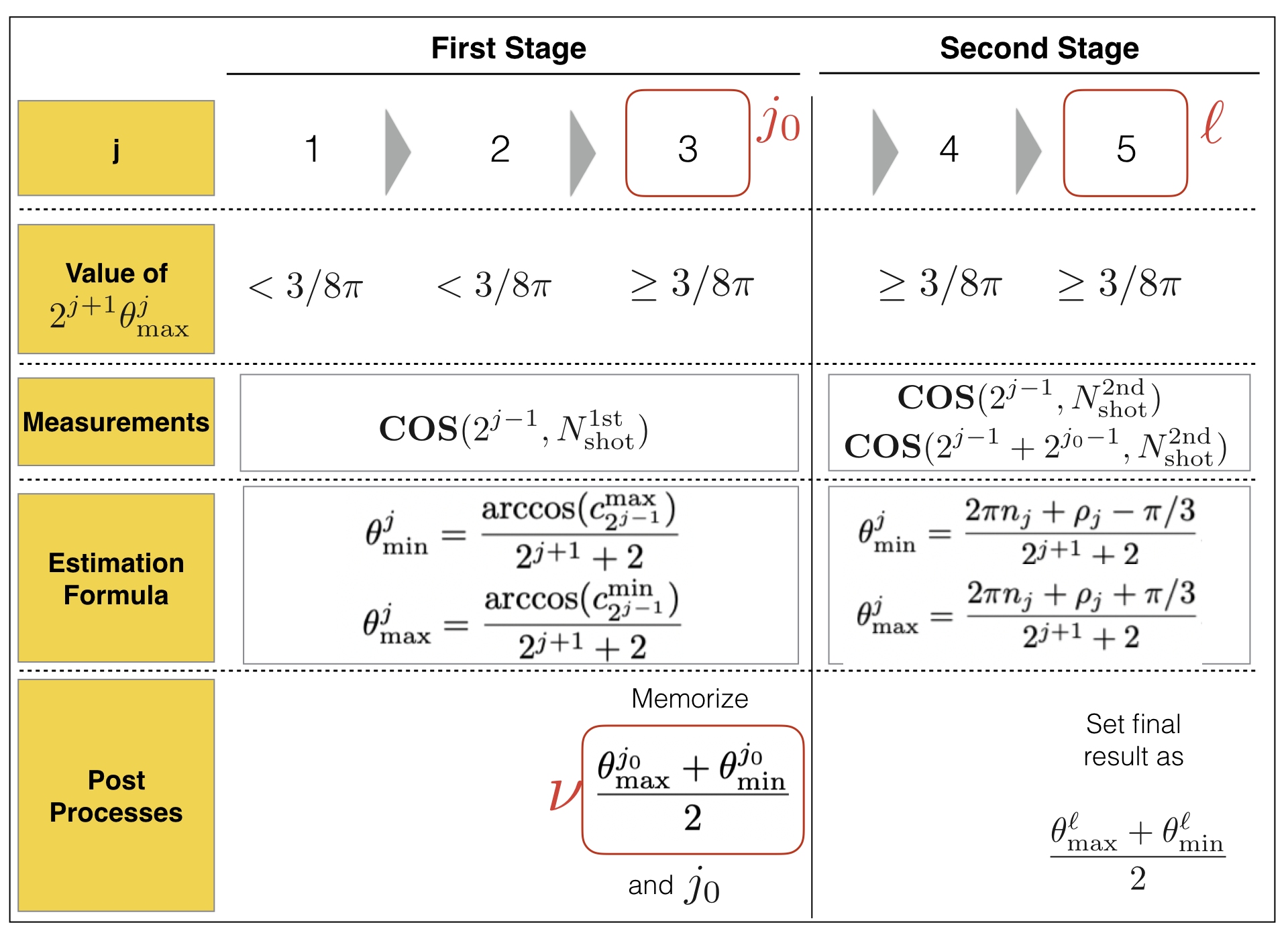}{The overview of our algorithm when $\ell=5$ and $j_0=3$.}{summary-of-algorithm}
\subsection{Complexity Upper Bound}
\label{subsection-complexity}

As we show in Appendix, by using our proposed algorithm, the required query complexity ($N_{\rm orac}$) with which the estimation error of $\amp$ is less than $\epsilon$ with the probability less than $\delta$ is bounded as
\begin{flalign}
\qquad
\label{complexity-estimation}
N_{\rm orac} < \frac{4.1 \cdot 10^3}{\epsilon} \ln\left(\frac{4\log_2(2\pi/3\epsilon)}{\delta}\right). 
\end{flalign}
The worst case is realized when the algorithm moves to the second stage at the first iteration (when $j=1$). We see that the upper bound of $N_{\rm orac}$ almost achieves Heisenberg scaling: ($N_{\rm orac} \propto 1/\epsilon$) because the dependency of the factor $\ln(\log_2(\pi/\epsilon))$ on $\epsilon$ is small, e.g.,  even when $\epsilon = 10^{-20}$, the factor is at most $6$. The tightest upper bound in previous literature is give by \cite{iterative-amplitude-estimation} as 
$N_{\rm orac} < \frac{1.15 \cdot 10^6}{\epsilon}\ln\left(\frac{2}{\delta} \log_3 \left(\frac{3\pi}{20\epsilon}\right)\right)$
in our notation. We see that the constant factor is $O(10^2)$ times smaller in our algorithm.

Although detail discussion is made in Appendix, here we briefly show why the upper bound is proportional to $1/\epsilon$. In order for $\epsilon$ to be bounded as (\ref{amp-error}), it is suffice that the errors of all $\estc{}{2^{j-1}}$s used in our algorithm are less than $1/20$, which is realized if $N_{\rm shot} \sim O(1000\log\left(1/\delta\right))$ measurements for each $j$. The number of oracle call in each $j$ is about $2^{j - 1}$ for each measurement. Thus, $N_{\rm orac} \sim N_{\rm shot}\sum_{j=1}^{j=\ell} 2^{j-1} = N_{\rm shot} 2^{\ell} \propto N_{\rm shot}/\epsilon$ as we expected.
\figmedium{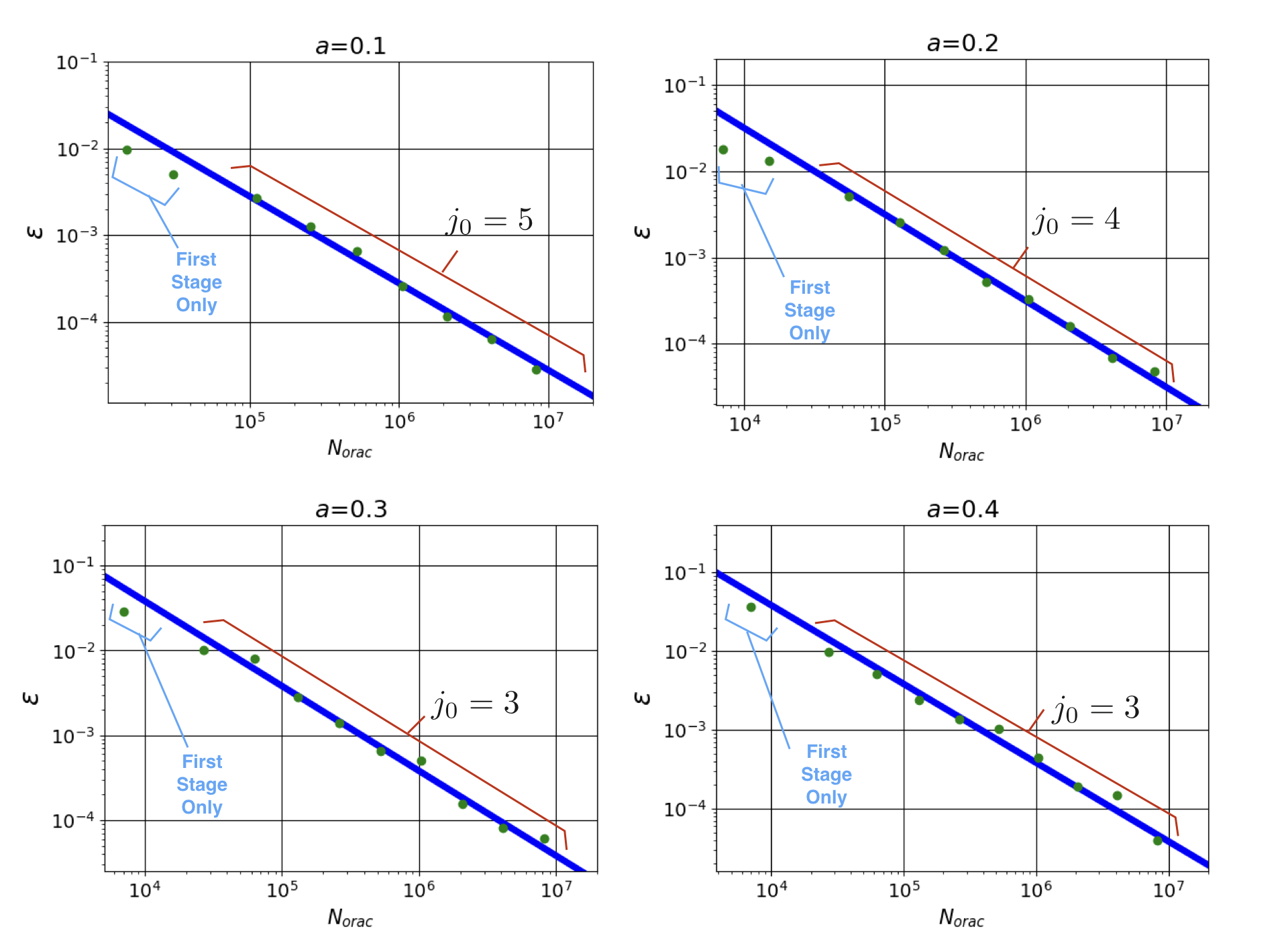}{Estimation error $\epsilon$ vs $N_{\rm orac}$ for $\amp=0.1$(left top), $\amp=0.2$(right top), $\amp=0.3$(left bottom) and $\amp=0.4$(right bottom). The green dots are plotted so that the estimation errors in 1000 trials are equals to or smaller than the plotted value. The green dots are fitted with $\log_{10}N_{\rm orac} = -\log_{10}(\epsilon) + b$ and shown as blue lines. The value of $j_0$ is also shown for each data point. }{results}

\section{Numerical Experiment}
\label{experiment}
In this section, we verify the validity of the algorithm introduced in Section~\ref{algorithm} by numerical experiments. We choose $\amp = 0.1, 0.2, 0.3,$ and $0.4$ as the amplitudes estimated. 
$\delta_c$ is taken as $0.01$.
We compute $N_{\rm orac}$ and the estimation error $\epsilon$ with changing the total number of algorithm steps $\ell$. In each parameter set $(\amp,\ell)$, we execute 1000 trials of the algorithm.

The computation results are shown in Fig.~\ref{results}. For each $N_{\rm orac}$, we plot the estimation errors(green dots) so that $95\%$ of the estimation errors in 1000 trials are equals to or smaller than the plotted value. In the same figure, we also show $j_0$. For data points where the algorithm does not go to the second stage, we write ``First Stage Only" instead of writing the value of $j_0$. The data points are fitted with $\log_{10}(N_{\rm orac}) = -\log_{10}(\epsilon) + b$ (blue lines) where the fitting parameter $b$ is determined by the least-squares.

Here is the list of notable points:
\begin{itemize}
\item As expected, the Heisenberg scaling $N_{\rm orac} \leq C \times 1/\epsilon$ is almost achieved.
\item In ``First Stage Only" cases, $\epsilon$ tends to be below the blue line, i.e., required $N_{\rm orac}$ is small for fixed $1/\epsilon$ compared with the case when the algorithm goes to the second stage. Because in ``First Stage Only'' cases, the cause of the error is limited; only $\cos(2^{j+1}+2)\theta$ is needed to estimate $2^{j+1}\theta$.
\item As $\amp$ increases, $j_0$ decreases as long as the algorithm goes to the second stage. Because, as $\amp$ increases, $2^{j+1}\theta_{\rm max}\geq3/8\pi$ is satisfied with smaller $j$.
\end{itemize}

\section{Conclusion}
The quantum amplitude estimation is an important problem that can be applied in various applications. Recently, the way of solving the problem without the phase estimation has been studied. Some of them suggest algorithms which 
achieve Heisenberg scaling ($N_{\rm orac} \leq C \times 1/\epsilon$) and they give rigorous proof. However the constant factor $C$ in each algorithm is large. Our contribution in this paper is providing an algorithm which almost achieves Heisenberg scaling and the constant factor is smaller than previous methods. We also give proof of the upper bound. 

In a practical usage of the algorithm, some improvements might be possible. Although we determine the values of $\Nfirst$ and $\Nsecond$ at the beginning of the algorithm for simplicity, we can reduce those values by iteratively determining them. For example, in the second stage, $\Nsecond$ can be smaller than that in (\ref{n-formula}) as long as $n_j$ in (\ref{integer-mod-formula}) is uniquely determined. Investigating those possible improvements are left for future works.

The effect of noise should also be examined. Although our algorithm can reduce the depth of the circuit compared to the quantum phase estimation algorithm, the required depth is still $O(1/\epsilon)$ and the effect of noise is not neglectable. Thus, studying how to tailor noise in our algorithm would be important for discussing the practicability of our algorithm, which is also left for future works.

\section*{Acknowledgement}
We acknowledge Naoki Yamamoto for insightful discussions and constructive comments. We also thank the two anonymous reviewers whose comments/suggestions helped improve and clarify this manuscript.
\label{discussion}

\appendix
\section{Proof of Complexity Upper Bound}
\label{appendix-upper-bound}
In this appendix, we provide a proof of the complexity upper bound. 
\begin{th.}
\label{upper-bound-theorem}
The following upper bound holds for $N_{\rm orac}$:
\begin{flalign}
\label{upper-bound-theorem}
N_{\rm orac} < \frac{4.1\cdot 10^3}{\epsilon} \ln\left(\frac{4\log_2(2\pi/3\epsilon)}{\delta}\right).
\end{flalign}
\end{th.}

\begin{proof}
Our strategy to obtain the upper bound is calculating the required number of $\Nfirst$ and $\Nsecond$ for the algorithm to work correctly with the probability $1-\delta$. Both upper bounds of $\Nfirst$ and $\Nsecond$ can be derived from the condition that our algorithm works correctly in the second stage because even though the condition from the first stage also bounds $\Nfirst$ loosely, the most strict upper bound of $\Nfirst$ can be gotten from the condition that the estimation error of $\nu$ is small enough. Thus, in the following, we only discuss the condition from the second stage. 


In the second stage, as we mention in Section \ref{section-algorithm}, the algorithm works correctly as long as $\Delta \rho_j \leq \pi/3$. Even though the conditions for ${\rm atan}\left(\ests{}{2^{j-1}},\estc{}{2^{j-1}}\right)$ derived from $\Delta \rho_j \leq \pi/3$ are different depending on whether the confidence interval of $\rho_j$ is the connected confidence interval or the disconnected confidence interval, the required precisions for $\ests{}{2^{j-1}}$ and $\estc{}{2^{j-1}}$ do not change depending on the interval type. Therefore, in the following, we discuss only the case of the connected confidence interval.
Then, the condition $\Delta \rho_j \leq \pi/3$ can be converted to
\begin{flalign}
\label{rho-j-error-estimation-difficult}
\left|{\rm atan}\left(\ests{}{2^{j-1}},\estc{}{2^{j-1}}\right) - {\rm atan}\left(\ests{\ast}{2^{j-1}}, \estc{\ast}{2^{j-1}}\right)\right| \leq \frac{\pi}{3}
\end{flalign}
where $\ests{\ast}{2^{j-1}}, \estc{\ast}{2^{j-1}}$ are the true values of $\ests{}{2^{j-1}}$ and $\estc{}{2^{j-1}}$ respectively. Given $\Delta \estc{}{2^{j-1}} = |\estc{}{2^{j-1}} - \estc{\ast}{2^{j-1}}|$, $\Delta \ests{}{2^{j-1}} = |\ests{}{2^{j-1}} - \ests{\ast}{2^{j-1}}|$, from (\ref{atan-theorem}) in Appendix \ref{proof-of-atan-theorem}, the following inequality holds for the left hand side of (\ref{rho-j-error-estimation-difficult}):
\begin{flalign}
\label{atan-error}
\left|{\rm atan}\left(\ests{}{2^{j-1}},\estc{}{2^{j-1}}\right) - {\rm atan}\left(\ests{\ast}{2^{j-1}}, \estc{\ast}{2^{j-1}}\right)\right| 
<  \max(2\Delta\estc{}{2^{j-1}} + 2\Delta \ests{}{2^{j-1}}, 3\Delta\estc{}{2^{j-1}})
\end{flalign}
as long as $\Delta \estc{}{2^{j-1}} < 1/4	$ and $\Delta \ests{}{2^{j-1}} < 1/3$. On the other hand, from (\ref{p-sin-estimation}), it holds that
\begin{flalign}
\label{sin-error}
\Delta \ests{}{2^{j-1}} &= \left|\frac{-\ests{\ast}{2^{j-1}}(\sin \nu - \sin (\nu - \Delta\nu))
+ \estc{\ast}{2^{j-1}}(\cos\nu - \cos(\nu - \Delta\nu))
+ \Delta \estc{}{2^{j-1}}\cos\nu + \Delta\estc{}{2^{j-1} + 2^{j_0-1}}}{\sin\nu}\right| \\ 
&\leq \frac{\sqrt{2 - 2\cos(\Delta \nu)} + |\Delta\estc{}{2^{j-1}}\cos\nu| + |\Delta\estc{}{2^{j-1} + 2^{j_0-1}}|}{\sin\nu} 
\end{flalign}
where $\Delta \nu = \nu - 2^{j_0+1}\theta$ and $3\pi/8 - |\Delta \nu| \leq \nu \leq 3\pi/4 - |\Delta \nu|$.
Thus, if at least the estimation errors are bounded as
\begin{flalign}
\label{c-2-inequality}
\Delta \estc{}{2^{j-1}} &\leq \frac{1}{9} \\
\quad \Delta \estc{}{2^{j-1} + 2^{j_0-1}} &\leq \frac{1}{9}, \label{cjj0-inequality}\\
|\Delta\nu| &< \frac{\pi}{60} \label{nu-inequality}
\end{flalign}
then it holds
\begin{flalign}
\label{sin-error-estimation-easy}
\Delta \ests{}{2^{j-1}} < \frac{\sqrt{2 - 2\cos(\frac{\pi}{60})} + \frac{1}{9}|\cos(\frac{3\pi}{4} - \frac{\pi}{60})| + \frac{1}{9}}{\sin(\frac{3\pi}{4} - \frac{\pi}{60})} < \frac{1}{3}.
\end{flalign}
As a result, 
\begin{flalign}
\label{atan-error-result}
\left|{\rm atan}\left(\ests{}{2^{j-1}},\estc{}{2^{j-1}}\right) - {\rm atan}\left(\ests{\ast}{2^{j-1}}, \estc{\ast}{2^{j-1}}\right)\right| 
< \max(2\cdot \frac{1}{9} + 2\cdot\frac{1}{3}, 3\cdot \frac{1}{9}) < \frac{\pi}{3}
\end{flalign}
is satisfied. 
Thus, by using (\ref{chernoff-bound}), if
\begin{flalign}
\label{n_inequality}
\Nsecond = 972 \ln\left(\frac{2}{\delta_c}\right)
\end{flalign}
then both the conditions (\ref{c-2-inequality}) and (\ref{cjj0-inequality}) is satisfied with the probability $1-2\delta_c$.
On the other hand, (\ref{nu-inequality}) is achieved if at least
\begin{flalign}
\label{rho-j-error-estimation-condition}
\Delta \estc{}{2^{j-1}} < \frac{1}{9\sqrt{2}},
\end{flalign}
holds in the first stage because
\begin{flalign}
\Delta \nu &= \frac{1}{2} \left(\arccos\left(\estc{\min}{2^{j_0 - 1}}\right) - \arccos\left(\estc{\max}{2^{j_0} - 1}\right)\right) \nonumber \\
&< \frac{1}{2}\left(\arccos\left(\cos\left(\frac{3\pi}{4}\right)\right) - \arccos\left(\cos\left(\frac{3\pi}{4}\right)  + \frac{1}{9\sqrt{2}}\right)\right) \nonumber \\
		   &< \frac{\pi}{60}.
\end{flalign}
Thus, by using (\ref{chernoff-bound}) again, it is shown that if
\begin{flalign}
\label{n_2_inequality}
\Nfirst = 1944 \ln\left(\frac{2}{\delta_c}\right).
\end{flalign}
then (\ref{nu-inequality}) is satisfied with the probability $1-\delta_c$. In summary, as far as (\ref{n_inequality}) and (\ref{n_2_inequality}) are satisfied, for all $j(>j_0)$, $\Delta\rho_j\leq\pi/3$ holds and (\ref{no-mod-formula}) is satisfied as long as all the estimates of cosines are inside the confidence interval; the probability is $(1-\delta_c)^{j_0 + 2(\ell - j_0)}>1-(2\ell - j_0)\delta_c$.

Finally, we evaluate the query complexity in the worst case. The worst case is that the algorithm moves to the second stage at the first iteration($j=1$). In the case, the number of oracle call is
\begin{flalign}
\label{oracle-estimation}
N_{\rm orac} < \Nfirst + \sum_{j=2}^{\ell}(2\Nsecond \times 2^{j-1}) = 1944\ln\left(\frac{2}{\delta_c}\right) + 1944(2^{\ell} - 2)\ln\left(\frac{2}{\delta_c}\right).
\end{flalign}
and the success probability of the algorithm is $1 - (2\ell -1)\delta_c$. Thus, if we demand that the success probability is more than $1 - \delta$ then $\delta_c < \delta/2\ell$ and 
\begin{flalign}
\label{difficult-condition}
N_{\rm orac} < 1944 \cdot 2^{\ell} \ln\left(\frac{4\ell}{\delta}\right).
\end{flalign}
By combining with (\ref{amp-error})
\begin{flalign}
\qquad
\label{final-bound}
&N_{\rm orac} < \frac{4.1\cdot 10^3}{\epsilon} \ln\left(\frac{4\log_2(2\pi/3\epsilon)}{\delta}\right). 
\end{flalign}
\end{proof}

\section{Theorem for atan function}
\label{proof-of-atan-theorem}
\begin{th.}
\label{atan-theorem}
When $c,c^{*},s \in [-1,1] $, $s^{*}$ takes one of the value of $\pm\sqrt{1 - c^{\ast 2}}$, $\Delta c = |c - c^{\ast}|$ and $\Delta s = |s - s^{\ast}|$, the following inequality holds:
\begin{flalign}
\label{atan-theorem}
|{\rm atan}(s, c) - {\rm atan}(s^{\ast}, c^{\ast})| < {\rm max}(2\Delta c + 2\Delta s, 3\Delta c)
\end{flalign}
if $\Delta s < 1/2$ and $\Delta c < 1/4$ and if there is no discontinuity of ${\rm atan}(s, c)$ in the intervals: $s^{\ast} - \Delta s \leq s \leq s^{\ast} + \Delta s $ and $c^{\ast} - \Delta c \leq c \leq c^{\ast} + \Delta c $.
\end{th.}
\begin{proof}
It is suffice to prove in following three cases: (i) $cc^{\ast} > 0$ (ii) $cc^{\ast} < 0$ and (iii) $cc^{\ast} = 0$. In case (i) $cc^{\ast} > 0$, using trigonometric addition formulas for $\arctan$, it holds that 
\begin{flalign}
\label{first-case}
|{\rm atan}(s, c) - {\rm atan}(s^{\ast}, c^{\ast})| &= |{\rm arctan}(s, c) - {\rm arctan}(s^{\ast}, c^{\ast})| \nonumber\\ 
&=\left|\arctan\left(\frac{s^{\ast} \Delta c - c^{\ast}\Delta s}{1 + c^{\ast}\Delta c + s^{\ast}\Delta s}\right)\right| \nonumber\\
 &\leq \left| \frac{|s^{\ast}|\Delta c + |c^{\ast}|\Delta s}{1 - |c^{\ast}|\Delta c - |s^{\ast}|\Delta s}\right| \nonumber\\
 &< 2 \Delta c + 2 \Delta s .
\end{flalign}
To show the last inequality, we use $1 - |c^{\ast}|\Delta c - |s^{\ast}|\Delta s > 1 - \sqrt{(1/3)^2 + (1/4)^2}> 1/2$ . 

In case (ii) $cc^{\ast} < 0$,
\begin{flalign}
\label{second-case}
|{\rm atan}(s, c) - {\rm atan}(s^{\ast}, c^{\ast})| 
&= \lim_{\eta \to 0} 
\left(\left|\arctan(s, c) - \arctan(s, \eta)\right| \right. \nonumber \\
&\qquad + \left. \left|\arctan(s, \eta) - \arctan(s^{\ast}, -\eta)\right| \right) \nonumber \\
&\qquad + \left. \left|\arctan(s^{\ast}, -\eta) - \arctan(s^{\ast}, c^{\ast})\right| \right)
\end{flalign}
where the sign of $\eta$ is same as that of $c$. The first term in (\ref{second-case}) can be bounded as
\begin{flalign}
\label{mean-value-first}
\lim_{\eta \to 0} \left|\arctan(s, c) - \arctan(s, \eta)\right|  
&= \lim_{\eta \to 0} \left|\frac{\partial}{\partial c}\arctan\left(\frac{s}{c}\right)|_{c = c_0} (c - \eta) \right| \nonumber \\
&= \lim_{\eta \to 0} \left|\frac{-s}{c_0^2 + s^2} (c - \eta)\right| \nonumber \\
&\leq \left|\frac{s}{c_0^2 + s^2} c\right| \nonumber \\
&\leq \left| \frac{1}{(c_{\ast} - (c_{\ast} - c_0))^2 + (s_{\ast} - (s_{\ast} - s))^2} c\right| \nonumber \\
&\leq \left| \frac{1}{\left(\frac{3}{5} - \frac{1}{4}\right)^2 + \left(\frac{4}{5} - \frac{1}{3}\right)^2} c\right| \nonumber \\
& < 3|c|
\end{flalign}
where $c_0$ take the value between $\eta$ and $c$, and we use the mean value theorem for showing the first equality. Similary, 
\begin{flalign}
\label{mean-value-last}
\lim_{\eta \to 0} \left|\arctan(s_{\ast}, -\eta) - \arctan(s_{\ast}, c_{\ast})\right|  
&< 3|c_{\ast}|.
\end{flalign}
By substituting (\ref{mean-value-first}), (\ref{mean-value-last}) and $\lim_{\eta \to 0} \left|\arctan(s, \eta) - \arctan(s^{\ast}, -\eta)\right| = 0$ (that follows from no-discontinuity condition)
to the right-hand side of (\ref{second-case}), it follows
\begin{flalign}
\label{second-case-result}
|{\rm atan}(s, c) - {\rm atan}(s^{\ast}, c^{\ast})| < 3 (|c| + |c_{\ast}|) = 3 \Delta c.
\end{flalign}
The last equality holds because the signs of $c$ and $c_{\ast}$ are different.

In case (iii) $cc_{\ast} = 0$, when $c_{\ast} = 0$,  
\begin{flalign}
\label{third-case}
|{\rm atan}(s, c) - {\rm atan}(s^{\ast}, c^{\ast})| 
&= \lim_{\eta \to 0} 
\left(\left|\pm \frac{\pi}{2} - \arctan(s_{\ast}, \eta)\right| + \left|\arctan(s_{\ast}, \eta) - \arctan(s, \eta)\right| \nonumber \right. \nonumber\\
&\qquad \qquad + \left. \left|\arctan(s, \eta) - \arctan(s, c)\right| \right)
\end{flalign}
where the sign $\pm$ is the same as the sign of $s$ and the sign of $\eta$ is same as that of $c$. The values of the first line go to $0$ and the value of the second line can be evaluated by the same arguments as (\ref{mean-value-first}). Thus, it follows 
\begin{flalign}
\label{third-case-result-1}
|{\rm atan}(s, c) - {\rm atan}(s^{\ast}, c^{\ast})| < 3 (|c|) = 3 \Delta c.
\end{flalign}
By the same discussion, when $c = 0$
\begin{flalign}
\label{third-case-result-1}
|{\rm atan}(s, c) - {\rm atan}(s^{\ast}, c^{\ast})| < 3 (|c_{\ast}|) = 3 \Delta c.
\end{flalign}
In all of the three cases, (\ref{atan-theorem}) is proved.
\end{proof}
\end{document}